# Equilibrium adsorption and self-assembly of patchy colloids in microchannels

Bennett D. Marshall

*ExxonMobil Research and Engineering, 22777 Springwoods Village Parkway, Spring TX 77389*

**Abstract**

A new theory is developed to describe the equilibrium adsorption and self-assembly of patchy colloids in microchannels. The adsorption theory is developed in classical density functional theory, with the adsorbed phase and fluid phase chemical potentials modelled using thermodynamic perturbation theory. Adsorption of non – patchy colloids in microchannels is typically achieved through non – equilibrium routes such as spin coating and evaporation. These methods are required due to the entropic penalty of adsorption. In this work we propose that the introduction of patches on the colloids greatly enhances the temperature dependent and reversible adsorption of colloids in microchannels. It is shown how bulk fluid density, temperature, patch size and channel diameter can be manipulated to achieve the adsorption and self-assembly of patchy colloids in microchannels.

Bennettd1980@gmail.com

# I: Introduction

The development of smart, controlled and functional materials through the reversible self-assembly of microscopic building blocks is a cornerstone of soft matter science. How can the forces between these microscopic building blocks be controlled to achieve a desired material? In recent years patchy colloids[1,2] have emerged as a platform for this type of assembly. Patchy colloids are colloids which contain discrete attractive patches[3]. The patches result in an anisotropy of the potential of interaction between colloids. The number[4], size[5] and arrangement[6] of patches can be controlled to achieve a desired self – assembled material. Specific experimental realizations include the self – assembly of patchy colloids into colloidal molecules [7,8] as well as low dimensional crystals such as the Kagome lattice[9]. In each case, the anisotropy of the patchy attractions is exploited to obtain a desired result.

The anisotropy and limited valence of the inter-colloid potential for patchy colloids shares many similarities to the intermolecular potential of hydrogen bonding fluids[10]. For this reason, statistical thermodynamic theories which describe the thermodynamics of hydrogen bonding fluids are often applied to patchy colloids[3]. Wertheim's thermodynamic perturbation theory TPT[11–14] has proven particularly valuable in the description of the self – assembly and phase behavior patchy colloid fluids.[15–20] Recently, Marshall[21] extended TPT to describe associating fluids in 1D spatial confinement, yet allowed to explore the full 3D orientation space. It was shown how the interplay between the anisotropy of the intermolecular potential and the 1D spatial confinement, resulted in enhanced association and an induced orientational order within clusters.

An area of interest in colloid science is the self-assembly of colloids (non-patchy) into microchannels[22]. This has application in colloidal lithography[23] and the self-assembly of micro - devices[24]. In the fluid phase, colloids will not strongly adsorb in these channels due to the confinement effect. To achieve appreciable adsorption, it is typical to use some external force such as spin coating[22] or evaporation[23] to force the colloids in the microchannels in a non-equilibrium way. Also, entropic depletion forces induced by considering colloid / polymer mixtures have been used to assemble colloids in microchannels.[25]

In this manuscript we ask the question, can the introduction of attractive patches on the colloids result in a temperature dependent and reversible self – assembly of the colloids in the microchannels? The philosophy being that the enhanced association observed in 1D associating fluids[21] may result in a preference for adsorption in these quasi – 1D microchannels. We study this system theoretically by applying classical density functional theory (DFT) to derive the density of colloids adsorbed in microchannels in equilibrium with a bulk fluid of patchy colloids. We restrict our attention to microchannels of diameters less than twice that of the patchy colloids. In this regime single file diffusion will hold.

To model the inter-colloid potentials, we employ the Kern – Frenkel[10] (KF) model of patchy colloids. The KF model is a primitive association model which represents the colloidal interactions as conical square well[26] association sites on a hard spherical core. This approach has been extensively used to numerically study the self – assembly of patchy colloid fluids.[4,5] To evaluate the required chemical potentials in the DFT, we employ Wertheim's thermodynamic perturbation theory (TPT) for associating fluids.[12–14] For the chemical potential of the bulk fluid phase we use the standard TPT1 treatment[26], while for the adsorbed phase we use the extension of TPT to 1D fluids of Marshall[21]. In what follows, it is shown how temperature, bulk fluid density, patch size and pore diameter can be manipulated to achieve strong adsorption of patchy colloids in microchannels.

## II: Theory

In this section we develop the theory for the adsorption and self-assembly of patchy colloids of diameter *d* in microchannels. Patchy colloids are modelled as hard spherical cores with short range directional association sites. The intermolecular potential between two colloids is given below

$$\phi(12) = \phi_{HS}(r_{12}) + \phi_{AS}(12) \tag{1}$$

where $(1) = (\vec{r}_1, \Omega_1)$ is short hand for the position $\vec{r}_1$ and orientation $\Omega_1$ of colloid 1 and $r_{12}$ is the distance between the two colloids. The hard sphere reference potential is

$$\phi_{HS}(r_{12}) = \begin{cases} \infty & r_{12} < d \\ 0 & otherwise \end{cases} \tag{2}$$

For the association interactions we use the two site model discussed in ref [21] which consist of type *A* and type *B* association sites located on opposite poles of the sphere in an axisymmetric arrangement.

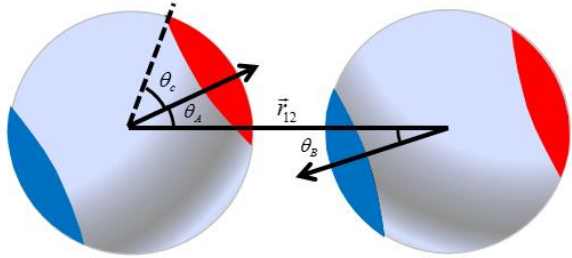

**Figure 1:** Representation of two patchy colloids interacting with the potential Eqns. (1) – (4). (Color online)

The *A* and *B* association sites are of differing functionality with *AB* attractions, but no *AA* or *BB* attractions. For this model the association potential is given as a sum over site attractions

$$\phi_{AS}(12) = \phi_{AB}(12) + \phi_{BA}(12) \tag{3}$$

The site – site potential $\phi_{AB}$ accounts for the attraction of site *A* on colloid 1 to site *B* on colloid 2. We model the site – site potentials using conical square well CSW association sites[26]

$$\phi_{AB}(12) = \begin{cases} -\varepsilon_{AB}, & r_{12} \leq r_c \text{ and } \theta_{A1} \leq \theta_c \text{ and } \theta_{B2} \leq \theta_c \\ 0 & \text{otherwise} \end{cases} \quad (4)$$

which states that if colloid 1 and 2 are within a distance $r_c$ of each other and each colloid is oriented such that the angles between the site orientation vectors and the vector connecting the two segments, $\theta_{A1}$ for colloid 1 and $\theta_{B2}$ for colloid 2, are both less than the critical angle $\theta_c$, the two sites are considered bonded and the energy of the system is decreased by a factor $\varepsilon_{AB}$. Figure 1 outlines two colloids interacting with CSW association sites. This is also known as the Kern – Frenkel model of patchy colloids.[10] Consistent with previous experimental results[7,8], we consider colloids with $r_c = 1.1d$

To develop the theory of adsorption we assume the microchannel length *L* is large, so edge effects are negligible. We restrict our attention to microchannels whose diameters $D_c$ obey the following relation (for the case of square channels $D_c$ represents channel width)

$$D_c < 2d \quad (5)$$

The restriction given by Eq. (5) prevents bypassing of colloids in the channel, preserving a given order, resulting in line diffusion. We derive the adsorption isotherm in the framework of classical density functional theory, where it is assumed that a bulk fluid phase of density $\rho_b$ is in equilibrium with the inhomogeneous adsorbed phase. The grand free energy is given by[27]

$$\Omega = A - \int \rho(\vec{r})(\mu_b - \Phi(\vec{r}))d\vec{r} \quad (6)$$

Where *A* is the intrinsic Helmholtz free energy, $\mu_b$ is the bulk fluid phase chemical potential, $\rho(\vec{r})$ is the density as a function of position, and $\Phi$ is the external field. Taking the functional derivative of (6) with respect to the total density we obtain the Euler – Lagrange equation

$$\frac{\delta \Omega}{\delta \rho(\vec{r})} = \frac{\delta A}{\delta \rho(\vec{r})} - \mu_b + \Phi(\vec{r}) = 0 \tag{7}$$

Within the channel, the external field is simply a confining field which keeps the patchy colloids within the channel.

We decompose both the free energy functional derivative and the bulk chemical potential into ideal and excess contributions

$$\frac{\delta A}{\delta \rho(\vec{r})} = k_b T \ln \rho(\vec{r}) + \frac{\delta A^{EX}}{\delta \rho(\vec{r})}; \quad \mu_b = k_b T \ln \rho_b + \mu_b^{EX} \tag{8}$$

where $A^{EX}$ is the excess Helmoltz free energy functional. Combining (7) – (8) and solving for the density

$$\rho(\vec{r}) = \rho_b \exp\left(\frac{\mu_b^{EX}}{k_b T} - \frac{\delta A^{EX}/k_b T}{\delta \rho(\vec{r})} - \frac{\Phi(\vec{r})}{k_b T}\right) \tag{9}$$

Since Eq. (5) assumes line diffusion, an important quantity is the 1D density $\rho_a$ which gives the number of colloids per length of channel. To obtain $\rho_a$ we integrate (9) over the area $\sigma$ of the channel (neglecting any axial dependence)

$$\rho_a = \rho_b \sigma_{ac} \exp\left(\frac{\mu_b^{EX}}{k_b T}\right)\left\langle \exp\left(-\frac{\delta A^{EX}/k_b T}{\delta \rho(\vec{r})}\right)\right\rangle \tag{10}$$

Where the notation $\langle x \rangle$ represents an average of the variable $x$ over the area accessible to the centers of the colloids $\sigma_{ac}$. For cylindrical and square channels (for example) the relevant areas are given by

*Cylindrical*          *Square*

$$\sigma = \frac{\pi}{4} D_c^2 \qquad \sigma = D_c^2 \tag{11}$$

$$\sigma_{ac} = \frac{\pi}{4}(D_c - d)^2 \qquad \sigma_{ac} = (D_c - d)^2$$

Lastly, the free energy functional derivative in (10) is approximated with the 1D excess chemical potential $\mu_a^{EX}$ evaluated at the density $\rho_a$

$$\rho_a = \sigma_{ac}\rho_b \exp\left(\frac{\mu_b^{EX} - \mu_a^{EX}}{k_b T}\right) \tag{12}$$

In the absence of association, Eq. (12) reduces to the theory of Glandt[28] for adsorbing hard spheres in an infinite single file cylindrical pore. As can be seen in Eq. (12), the theoretical calculations depend on the accessible channel area $\sigma_{ac}$ and not any specific channel geometry. That is, predictions of (12) will be the same for rectangular and cylindrical geometries as long as $\sigma_{ac}$ remains constant. Also, the derivation given above is for the case of bulk fluid phase in equilibrium with a single file adsorbed phase (pore). For the case of surface grooves in contact with a fluid phase, we assume that the chemical potential within the groove is that of a 1D phase with the accessible channel area $\sigma_{ac}$ of an enclosed pore with equal diameter. It has been shown for adsorption on surfaces, that treating the adsorbed phase as 2D allows for accurate theoretical prediction of adsorption isotherms.[29]

We split the 1D excess chemical potential into contributions due to hard sphere repulsions and association attractions

$$\mu_a^{EX} = \mu_a^{HS} + \mu_a^{AS} \tag{13}$$

For the hard sphere contribution, we use the exact chemical potential for a Tonks fluid of 1D hard rods. For the association contribution of the adsorbed phase, resulting from the site – site potentials Eq. (4), we employ the recently developed theory of Marshall[21] which gives the association contribution to the chemical potential for the adsorbed phase as

(14)

$$\frac{\mu_a^{AS}}{k_B T} = \cos\theta_c \bar{X}_o \ln\frac{\bar{X}_o}{2} + (1 - \bar{X}_o \cos\theta_c)\ln\left(\left[\frac{\bar{X}_o}{2} + \frac{1-\bar{X}_o}{2(1-\cos\theta_c)}\right]Y_A^2\right) - \rho_a(1 - \bar{X}_A)\frac{\partial \ln \Psi}{\partial \rho_a} - \ln\left(\frac{1}{2}\right)$$

The term $\bar{X}_A$ is the fraction of unbonded association sites which is given by the relation

$$1 = \bar{X}_A + f_{AB} \Psi \rho_a \bar{X}_A^2 \kappa_{AB} \qquad (15)$$

where $f_{AB} = exp(\varepsilon_{AB}/k_bT) - 1$ is the association Mayer function, $\Psi = 2 \int_d^{rc} g_{HS}(z) dz$ is the integral of the hard sphere reference correlation function over all allowed associated states, and $\kappa_{AB}$ is the probability that two spheres are oriented correctly for association

$$\kappa_{AB} = \frac{\left(1 - \frac{\bar{X}_o}{\bar{X}_A} \cos\theta_c\right)^2}{4} \qquad (16)$$

The second term on the right hand side of Eq. (15) is then interpreted as the fraction of bonded association sites. This fraction is given by the product of probabilities that two colloids are positioned $\rho_a \Psi$, oriented $\kappa_{AB}$ and have unbonded association sites $\bar{X}_A^2$ to allow for association. These probabilities are then weighted by $f_{AB}$ which is the Boltzmann factor – 1. The – 1 is to enforce that there is zero association when the association energy is identically zero.

The fraction of unbonded colloids $\bar{X}_o$ is given by

$$\frac{1}{\bar{X}_o} = \cos\theta_c + \frac{1 - \cos\theta_c}{Y_A^2} \qquad (17)$$

where $Y_A$ is the fraction of association sites which are unbonded on colloids which have an orientation $|\cos\theta| \geq |\cos\theta_c|$. The angle $\theta$ is the angle between the site $A$ orientation vector and the $z$ axis. See Fig. 2 for an illustration.

$$\frac{1}{Y_A} = 1 + \frac{1}{2}\left(\frac{1}{\bar{X}_A} - 1\right)\frac{1}{\sqrt{\kappa_{AB}}} \qquad (18)$$

Equations (15) – (18) provide a closed set of equations; upon their solution the chemical potential in Eq. (14) can be evaluated. This approach has been shown to accurately[21] predict the self – assembly of colloids which obey the association potential Eqns. (1) - (4) in 1D.

In the development of the 1D association theory it was assumed that the spheres were confined to a single spatial dimension; however, there was no restriction placed on orientation as the spheres were allowed to freely rotate. The orientation dependence of the intermolecular potential combined with the 1D spatial treatment results in large orientationally ordered clusters in strongly associating systems. We quantify this orientational order with the order parameter S,

$$S = \frac{3\overline{\cos^2\theta}-1}{2} \qquad (19)$$

The orientational average is given by

$$\overline{\cos^2\theta} = \frac{\bar{X}_o}{3} + \frac{(1-\bar{X}_o)}{3}\frac{(1-\cos^3\theta_c)}{1-\cos\theta_c} \qquad (20)$$

For systems with perfect orientational order $S = 1$, while for systems without orientational order $S = 0$. The last quantity needed to evaluate Eq. (12) is the bulk fluid phase excess chemical potential. For this quantity we use the standard Wertheim's first order perturbation theory [14,30].

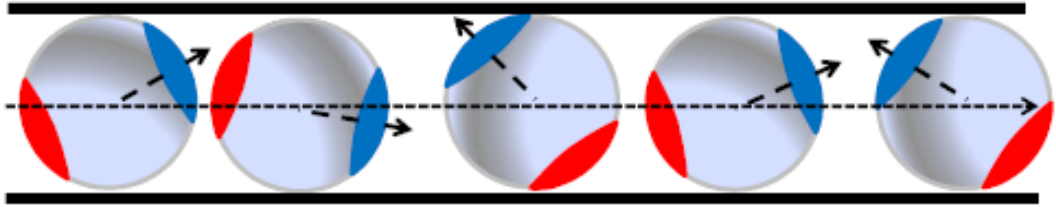

**Figure 2:** Patchy colloids interacting in a 1D channel. Short arrows represent orientation vector and long arrow represents z axis. Colloids are allowed to explore all Euler orientation angles (orientation vectors may point into or out of page). (Color online)

## III: Results

In this section we employ the theory developed in section II to calculate the adsorption and self-assembly of patchy colloids from a bulk fluid phase of density $\rho_b$ to a cylindrical microchannel of diameter $D_c$. An important quantity is the adsorption selectivity $K$, defined as the density in the pore divided by that in the bulk

$$K = \frac{\rho_a}{\sigma \rho_b} \tag{21}$$

Figure 3 shows the selectivity and associated cluster sizes (in both bulk and adsorbed phases) for a patchy colloid fluid with $\theta_c = 10°$ at reduced association energies $\varepsilon^* = \varepsilon_{AB} / k_b T = 0, 8, 10$ and 12 versus density in the bulk fluid phase in a channel with diameter $D_c = 1.5d$.

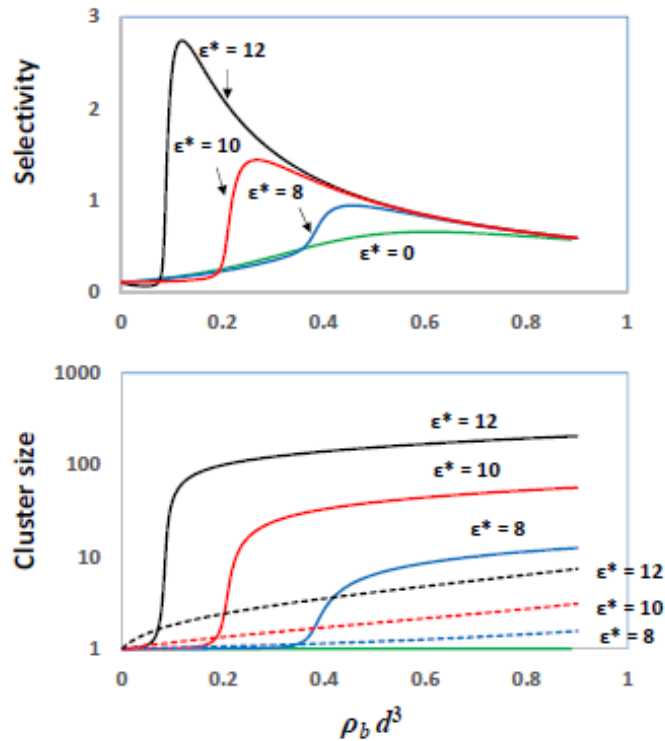

**Figure 3:** Theory results for patchy colloids with $\theta_c = 10°$ in a cylindrical channel of diameter $1.5d$, versus fluid phase density. Top: selectivity defined by Eq. (21). Bottom: Solid curves give average adsorbed phase cluster size, while dashed curves give fluid phase associated cluster size. (Color online)

For the hard sphere case $\varepsilon^* = 0$, the fluid is depleted from the channel at all fluid phase densities. The depletion is the result of an entropic penalty due to confinement in the channel, this is captured in the model through the area $\sigma_{ac}$. As bulk density is increased, the selectivity reaches a maximum at bulk densities near 0.6 and then decreases. This maximum and tail result from the inefficient 3D packing in the single file microchannel, and are simply the result of the selectivity definition Eq. (21).

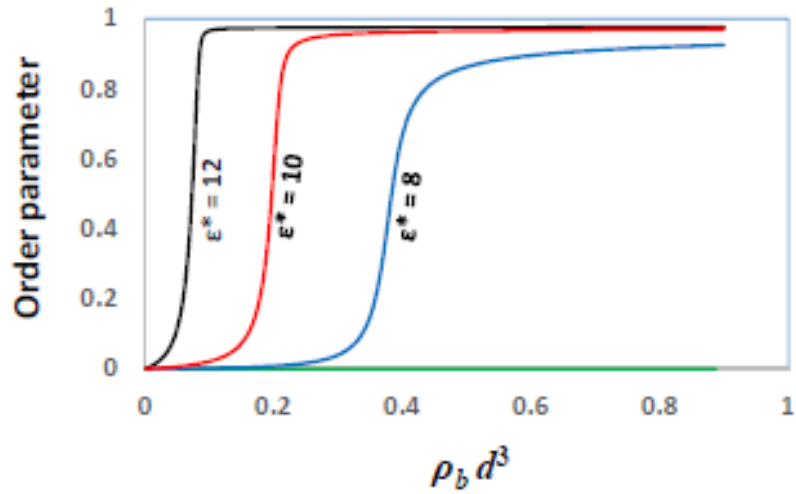

**Figure 4:** Adsorbed phase order parameters for same system as Fig. 3. (Color online)

Now, we turn our attention to the strongly associating (low T) case $\varepsilon^* = 12$. For low fluid phase densities ($< 0.1$) we see that association decreases selectivity as compared to the hard sphere case, but at a fluid density near 0.1 there is a drastic jump in selectivity as density is increased. The maximum selectivity obtained is near $K = 2.7$. This enhanced selectivity is the result of the anisotropy of the association interaction, as well as the single file confinement in the channel. A critical angle of 10° results in a large penalty in orientational entropy due to association, making it difficult to form large associated clusters in the fluid phase. However, since the adsorbed phase is nearly 1D, once a colloid has formed an association bond, it has a restricted number of orientations which results in a much lower entropic penalty in forming an additional association bond. For instance, at $\theta_c = 10°$ the probability two bonded colloids are oriented correctly to form a second association bond to each other in the bulk fluid is $\sim 5.8 \times 10^{-5}$, while in the adsorbed 1D fluid the probability is 0.25. This can be seen in the bottom panel of Fig. 3 which shows averaged adsorbed phase and fluid phase associated cluster sizes, as well as Fig. 4 which shows the corresponding adsorbed phase order parameters.

For fluid densities < 0.1 (at $\varepsilon^* = 12$) the adsorbed phase lacks orientational order and fluid phase clusters are larger than adsorbed phase. Near this point there is a rapid increase in adsorbed phase cluster size, which results in orientationally ordered associated clusters. This is the genesis of the observed jump in selectivity. For the weaker association (higher $T$) cases $\varepsilon^* = 10, 8$ we see similar behavior as $\varepsilon^* = 12$; however, the transition becomes less pronounced, more gradual and shifts to higher densities as temperature is increased. Note, this is not a true phase transition, but a polymerization transition as studied previously in 3D systems.[16]

The results in Figs. 3 and 4 are for a fixed channel diameter of $D_c = 1.5d$. It is also desired to determine the interplay of channel diameter and patch size on the self – assembly of patchy colloids in microchannels. Figure 5 displays the selectivity and average cluster sizes for the case of an adsorbing fluid ($\rho_b = 0.3d^3$ and $\varepsilon^* = 12$) versus channel diameter for patch sizes of $\theta_c = 0°, 5°, 20°$ and $30°$. For the hard sphere case ($\theta_c = 0°$) we compare to the Monte Carlo simulation results of Post and Kofke[31]. Initially, the hard sphere case shows a monotonic increase in selectivity with increasing pore diameter; this is followed by a maximum in selectivity near $D_c = 1.7d$, and then a decrease in selectivity until the single file limit of $D_c = 2d$. The model and simulation are in excellent agreement for $D_c < 1.7d$, and reasonable agreement at larger channel diameters. Increasing the patch size to $\theta_c = 5°$, we see a similar monotonic increase in selectivity for small channel sizes; however, near $D_c = 1.2d$, there is a sudden jump in selectivity reaching a maximum near $K = 1.8$. That represents an order of magnitude increase in selectivity as compared to the hard sphere case.

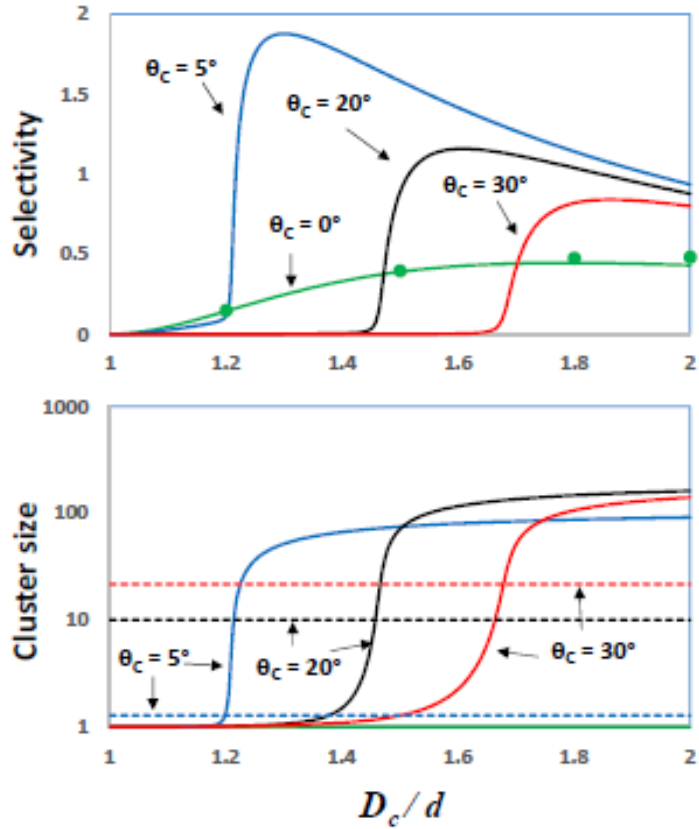

**Figure 5:** Theory results for patchy colloids at a reduced association energy $\varepsilon^* = 12$ and bulk fluid phase density $\rho_b = 0.3d^3$ versus cylindrical channel diameter at various patch sizes. Top: selectivity defined by Eq. (21). Symbols represent Monte Carlo simulation results for hard spheres[31]. Bottom: Solid curves give average adsorbed phase cluster size, while dashed curves give fluid phase associated cluster size. (Color online)

As in figure 3, this large increase in selectivity is a result of strong association in the adsorbed

phase and minimal association in the bulk fluid phase. This can be seen in the bottom panel of

Fig. 5. The bulk fluid phase average cluster size is of course unaffected by channel diameter and remains constant at 1.28. The average adsorbed phase cluster size remains close to 1 before the transition at $D_c = 1.2d$, and increases rapidly after this transition. Again, this is a result of the strong anisotropy of the association potential Eq. (4) for small patch sizes. Increasing the patch sizes to $\theta_c = 20°$ and $30°$ results in increased association in the bulk fluid phase. This results in a depletion of colloids from the channels for small $D_c$ as compared to the smaller patch cases. Eventually, as pore diameter is increased, there is a point where selectivity begins to increase; however, at larger patch sizes the transition is more gradual and less pronounced.

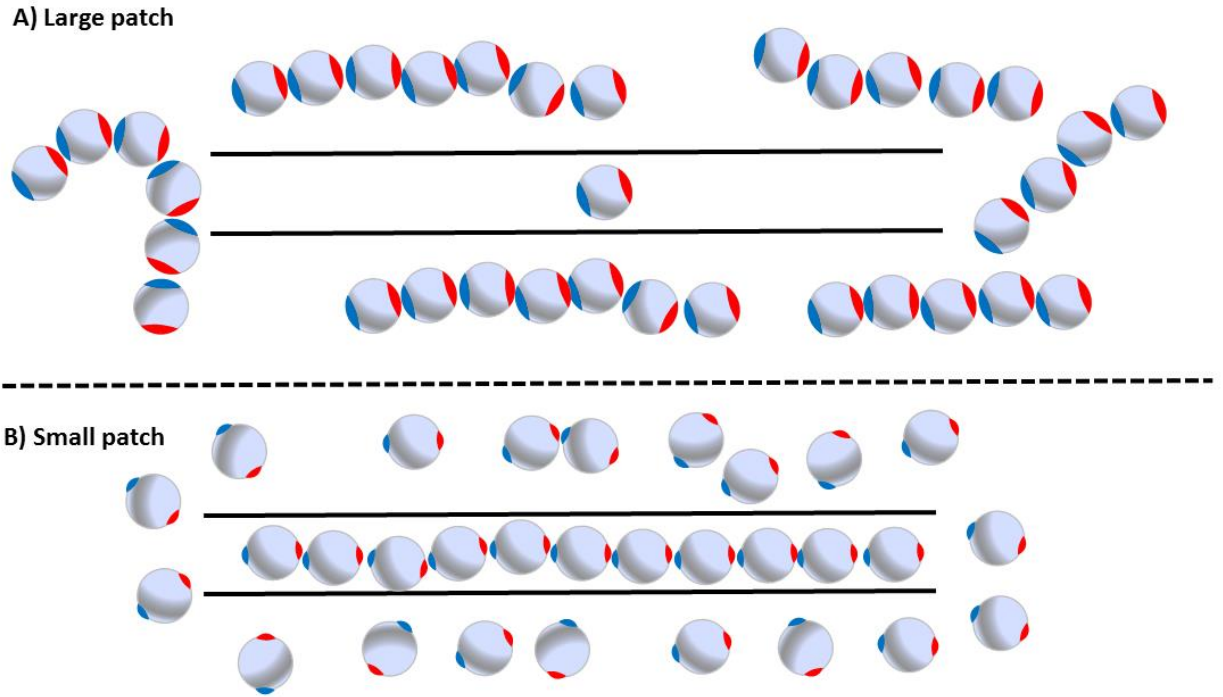

**Figure 6:** Diagram of large patch (top) and small patch (bottom) patchy colloid fluids interacting with a cylindrical channel (parallel lines). Large patch colloids remain depleted in the channel due to strong association in the bulk, while small patch colloids adsorb in the channel due to weak association in the bulk. (Color online)

**IV: Conclusions**

The results presented above can be summarized as follows. Self-assembly of patchy colloids in microchannels is a competition between the decrease in energy due to association, the translational entropic penalties due to confinement in the channel, and finally the penalty in decreased orientational entropy due to association. The maximum selectivity of patchy colloids in microchannels is found at low temperatures (which favor association) and small patch and channel sizes. At these small patch sizes the entropic penalty for association is large. However, in the channel this entropic penalty is lower due to the fact that once a colloid has formed one association bond, the probability it is oriented correctly to form a second association bond is enhanced as compared to the bulk fluid phase. The difference in entropic penalties due to association between the channel and bulk may result in a large increase in selectivity as compared to the hard sphere case, or even cases with different patch sizes. This is depicted pictorially in Fig. 6.

A major assumption used in the development of the theory is that the adsorbed phase free energy functional derivative can be approximated with a 1D chemical potential. For single file diffusion, this seems like a reasonable approximation. Indeed, Fig. 5 demonstrates the utility of this approximation in the description of hard sphere fluids. However, for patchy colloids which obey Eqns. (3) – (4), a comparison of theory predictions to "exact" numerical results obtained through grand canonical Monte Carlo simulations would be a point for future research.

Additionally, the analysis presented here is purely thermodynamic and does not address the issue of kinetics. How long would it take for the system to reach its equilibrium state? To obtain the final equilibrium structure the colloids need to diffuse through the channel and then, through rotational diffusion, orient for association. This would be an interesting case study using kinetic Monte Carlo.[32] An alternative realization of this system, which would limit frustration in the adsorbed phase, would be to choose a colloidal system with two identical self-associating patches. This would remove a possible kinetic barrier introduced by have left handed and right handed clusters for the bi – functional case studied here.

**Acknowledgments:**

The author would like to thank Lang Feng and Walter Chapman for proof reading of the manuscript